# Three-Dimensional Bubbly Flow Measurement Using Event-based Vision Sensor Cameras

A. Al Brahim[1,*], K. Rajamanickam[2], B. Lecordier[3], A. Taylor[1], Y. Hardalupas[1]

1: Department of Mechanical Engineering, Imperial College London, United Kingdom
2: Department of Mechanical and Aerospace Engineering, The University of Manchester, United Kingdom
3: INSA Rouen Normandie, Univ Rouen Normandie, CNRS, Normandie Univ, CORIA UMR 6614, France
* Correspondent author: a.al-brahim23@imperial.ac.uk

ABSTRACT

A three-camera Event-Based Vision Sensor (EVS) system is employed to perform three-dimensional measurements of bubble motion, morphology, and bubble–bubble interactions. The multi-EVS configuration mitigates the absence of direct depth information inherent to binary event-based imaging while preserving key advantages, including high temporal resolution, low latency, and reduced data throughput. The experimental configuration consisted of an octagonal tank equipped with a controlled particle release mechanism and an air diffuser. Camera synchronization and pulsed LED illumination were achieved using a dedicated signal generator and driver circuitry, while calibration was performed using pulsed-illumination recordings of a target acquired at multiple depths. A comprehensive, in-house computational framework was developed to process the event data for three-dimensional motion trajectory reconstruction. The validation of the developed tracking framework followed a rigorous multi-stage pipeline to ensure reconstruction fidelity. The framework was initially benchmarked against synthetic rendering cases of increasing kinematic complexity to evaluate 3D trajectory reconstruction accuracy under controlled conditions, achieving sub-millimeter global accuracy with root-mean-square error (RMSE) values ranging from 0.015 to 0.36 mm. Following numerical validation, physical baseline experiments were conducted using precisely manufactured particle releases through both a gated chamber and a single-particle claw opening mechanism. Subsequently, dynamic bubble plumes generated via multiple inlets across various compressed air flow rates were evaluated. The EVS framework successfully resolved dense bubble-bubble interactions, producing smooth, physically consistent three-dimensional trajectories across all tested conditions. Quantitative results showed strong agreement, yielding an overall velocity consistency exceeding 97% between conventional centroid tracking and independent velocity estimates derived from continuous-illumination event streaks. Overall, this methodology demonstrates a robust framework for high-speed volumetric tracking and bubble flow measurement. However, event oversaturation remains a key hardware limitation in ultra-dense regimes, as oversaturation in a single camera can cause system desynchronization.

## 1. Introduction

Bubbly flows play a critical role in numerous industrial and environmental processes. However, their characterization remains challenging due to the rapid evolution of bubble motion, shape, and interactions (Garbin et al., 2025). Since these bubbles undergo three-dimensional motion and morphological changes in space, various studies have utilized multi-camera imaging to investigate them. These systems typically capture bubble shadowgraphs of the bubble against a backlight illumination. Fu and Liu (Fu & Liu, 2018), and She et al. (She et al., 2021) used four synchronized cameras to investigate the 3D trajectory of a rising bubble and its induced flow field, whereas Wang et al. (H. Wang et al., 2025) examined bubble pair interactions and the resulting flow dynamics. Similarly, Di Nunno et al. (Di Nunno et al., 2020) have employed four synchronized cameras and the space carving technique to track the time evolution of kinematic and geometric properties of rising bubbles and a plunging jet, benchmarked against rigid spheres. Bubble swarms have also been investigated in the work of Hessenkemper et al. (Hessenkemper et al., 2022), Tan et al. (Tan et al., 2023), and Wang et al. (L. Wang et al., 2025).

Traditionally, three-dimensional bubble imaging is performed using High-Speed (HS) CMOS cameras. However, their high data rates, limited recording duration, and typical high cost hinder their use in characterizing transient bubbly flows, such as flow boiling (Lu et al., 2024). In this study, we investigate the use of a multi-camera Event-Based Vision Sensor (EVS) system as an alternative that overcomes these limitations. In contrast to conventional cameras, each pixel of the EVS camera sensor responds asynchronously to changes in the light intensity. A positive event is triggered when the increase in intensity exceeds a predefined positive threshold, while a negative event is generated when the intensity decreases below a predefined negative threshold.

Although EVS cameras have been used in a wide range of applications (Gallego et al., 2019), their use in multi-camera configurations has been predominantly limited to robotic applications (Ghosh & Gallego, 2025). Notable exceptions include the works of Wang et al. (Y. Wang et al., 2020) on particle tracking for flow reconstruction using stereo EVS, and Willert and Klinner (Willert & Klinner, 2025) on three-dimensional particle tracking using three EVS cameras for wall shear stress estimation.

The performance of EVS cameras for capturing two-dimensional bubbly flow was evaluated in our previous work (Al Brahim et al., n.d.) by comparing them with conventional HS camera imaging. We observed strong agreement between size and velocity measurements obtained simultaneously from both cameras. However, we also identified limitations in the EVS camera data. These limitations include event loss due to sensor oversaturation at high event rates, driven by rapid bubble motion and high bubble generation frequency. In addition, unlike the conventional camera, the EVS camera lacks depth information due to the inherently binary nature of event-based imaging.

In this study, we employ three synchronized EVS cameras to investigate three-dimensional bubble trajectories, shapes, and interactions of bubbles released from the bottom of an octagonal tank filled with a water–surfactant solution. The release of precisely manufactured, naturally buoyant particles is used to assess the tracking and calibration performance of the system. The accuracy of the multi-EVS camera measurements and their associated limitations are systematically evaluated to assess their suitability for three-dimensional bubbly flow characterization.

**2. Experimental Methodology**

The experimental setup is designed to observe three-dimensional trajectories of positively buoyant particles and bubbles in quiescent water under both pulsed and continuous illumination. The apparatus, illustrated in **Figure. 1** , is centered around an octagonal tank monitored by three event-based (EVS) cameras and illuminated by a trio of diffused Light-Emitting Diodes (LEDs) with dedicated drivers. A signal generator provides timing triggers, while a central computer powers and controls the EVS cameras and records synchronized signals.

The octagonal tank has side lengths of 60 mm and a height of 150 mm. Its base is configurable with either a particle chamber or an air diffuser. The particle chamber has three compartments in a triangular layout, with the primary compartment centered on the tank base. The two outer chambers are positioned at a direct center-to-center distance of 6 mm from this central point, each with a lateral X-axis offset of 3 mm. Each compartment measures 3.5 mm in diameter and 30 mm in depth. The compartments are designed to house polypropylene polymer particles (manufactured by Cospheric), with a diameter of 2.483 mm ± 0.004 mm and a density of 0.9 g/cc, making them positively buoyant in water. The difference between the particle diameters and the compartment diameter is designed to allow water to enter the chamber and enable the particles to float due to buoyancy, while remaining small enough to prevent the simultaneous release of multiple particles.

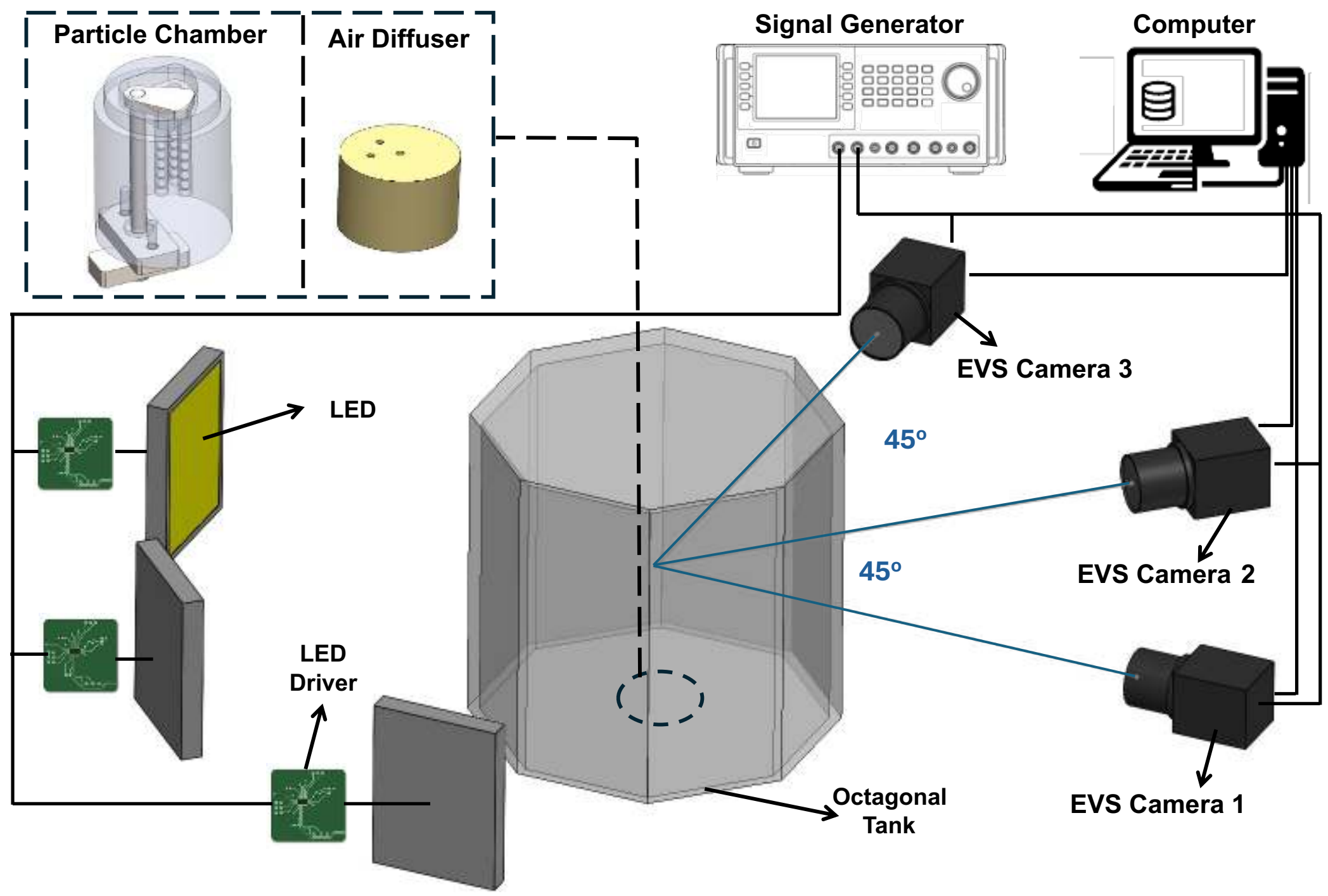

**Figure. 1** Experimental setup showing the octagonal tank and the three EVS cameras and corresponding LED illuminations.

During the loading process, the tank is drained, and particles are introduced individually using a syringe to prevent agglomeration. Subsequently, the compartments are filled to the brim and sealed with a motorized gate. The gate opening speed can be regulated to introduce controlled variability in particle release once the tank is fully filled. Although particles are ideally positioned at the center of each chamber compartment, they often come into contact with the walls, likely due to the dynamics of the initial water filling process and stacking within the compartment. This can result in slight variations of the initial conditions of the particles at the point of release.

For bubble generation, the tank is filled with a water/surfactant solution with surface tension of 27.7 mN/m. Compressed air is supplied at the base through a diffuser containing three 0.3 mm inlets: one central inlet and two additional inlets positioned 6 mm radially from the center. The supplied flow rates range from 10 to 20 mL/min. Both the airflow rate and surfactant concentration are adjusted to control bubble generation rate, motion, and size.

The utilized EVS cameras are the PROPHESEE EVK4 capable of recording at a resolution of 1280 x 720 pixels, with a pixel latency of less than 100 $\mu$s and a dynamic range above 120 dB. The cameras are positioned in a horizontal arc at -45°, 0°, and 45° relative to the central axis, positioned at a distance of 600 mm from the tank center. Each camera was equipped with an identical 50 mm macro lens set to an aperture of f/11 to ensure consistent imaging conditions across the system. Unlike frame-based cameras, EVS cameras’ settings can be adjusted before and after recording. Pre-recording adjustments include bias settings to control trigger sensitivity and the registration of positive and negative events, as well as the resolution of the region of interest (ROI) . The frame rate and event accumulation time window can be modified post-recording.

The LEDs can be operated in either continuous or pulsed modes. In the continuous configuration, they are connected directly to the power supply and activated prior to recording. Following the initiation of camera recording, a function generator (TTi TGP110) is triggered to provide a timing signal synchronized across all cameras. In pulsed mode, MOSFET-based LED drivers synchronously pulse all LEDs according to the timing and frequency specified by the function generator. In this mode, the EVS camera biases must be tuned to mitigate the generation of 'OFF' (negative) events caused by the rapid reduction in light intensity following each pulse.

To calibrate the cameras, a dot-pattern target consisting of 1 mm dots spaced 2 mm apart was captured at varying depths using all EVS cameras, as shown in **Figure. 2** (a). Since EVS cameras cannot detect static objects, the LED illumination was pulsed to make the target observable. Due to the stochastic characteristics of the EVS sensors, dot boundaries may exhibit noise when local intensity variations are insufficient to trigger events, as shown in **Figure. 2** (b). This effect is also apparent in the checkerboard pattern of **Figure. 2** (b), where inconsistencies in event detection cause some corners to appear connected, while others remain separated. Since precise identification of all pattern pixels is essential for accurate calibration, and checkerboard corners are more vulnerable to noise, the dot-pattern target was ultimately preferred.

Careful control of both illumination intensity and aperture setting is necessary to prevent light overexposure and underexposure. Overexposure can lead to blooming that diminishes the apparent size of the pattern features, while underexposure introduces temporal noise and may artificially inflate dot dimensions. To achieve optimal calibration quality, the LED source was covered with multiple diffusion layers to ensure uniform light intensity across the target images. The calibration images were processed using LaVision DaVis 8.4, yielding a calibration RMS fitting error of less than 0.2 under the pinhole camera model.

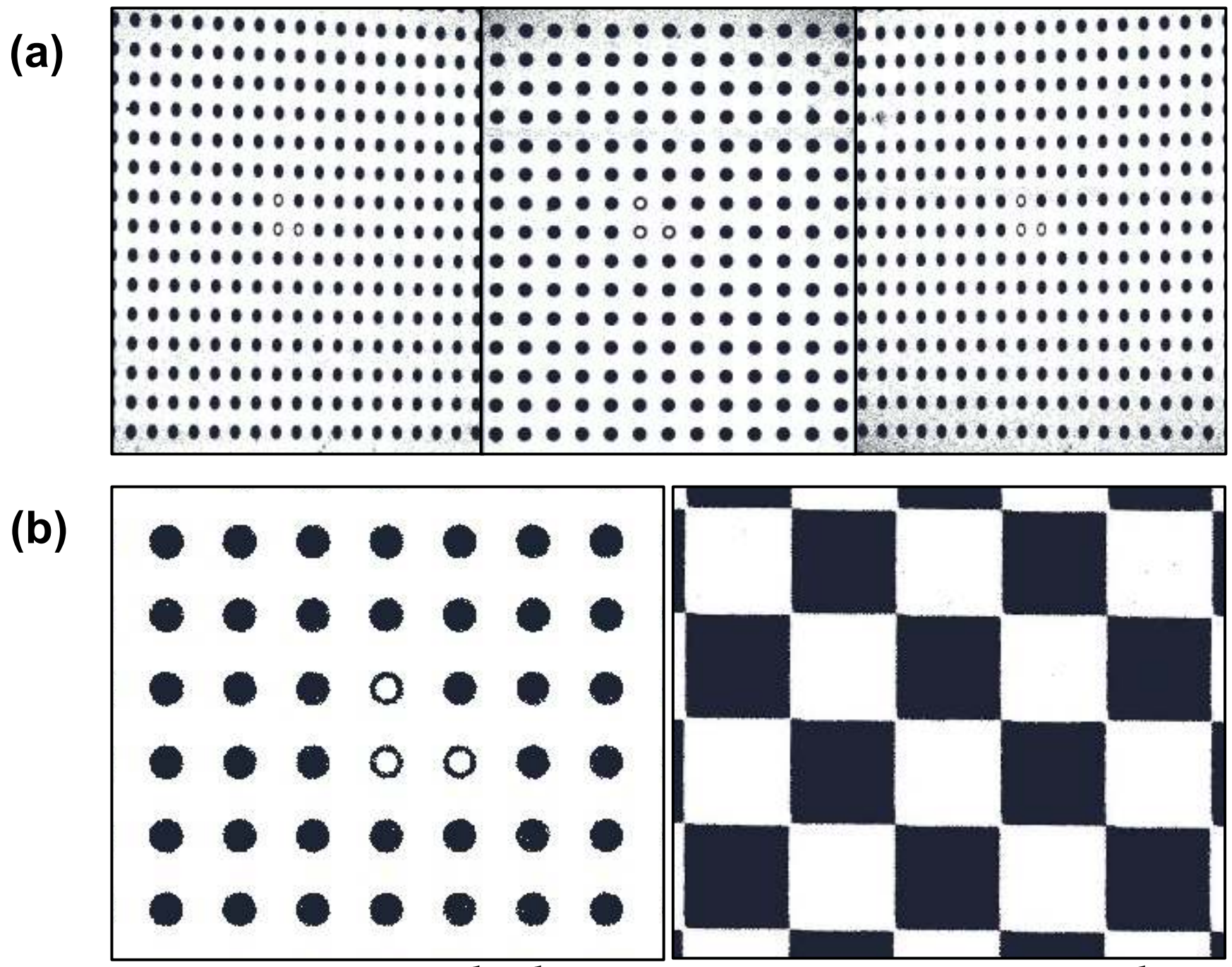

**Figure. 2** (a) Dot-pattern images acquired from the three EVS cameras for calibration at the target central position. (b) Comparison between dot and checkerboard calibration patterns captured under direct viewing from an EVS camera.

## 3. Image analysis and tracking algorithms

Following calibration, particle and bubble release experiments were recorded with the EVS cameras under both pulsed and continuous illumination conditions.

Under pulsed illumination, the reconstructed EVS frames resemble binary HS camera images, as can be observed in **Figure. 3**, since the sensor primarily responds to temporal changes in background illumination unobstructed by particle or bubble shadows. Each illumination pulse produces a substantial burst of detected events, which cannot be scanned and timestamped instantaneously by the EVS row arbiter. However, the temporal interval between successive pulses provides sufficient time for all generated events to be processed prior to the subsequent pulse. The EVS camera accumulation time is set to match the pulse frequency, allowing the cameras to construct frames that capture all events generated during each pulse. A more comprehensive discussion of the implementation and associated considerations for pulsed illumination is presented in Willert's work (Willert, 2023).

Under continuous illumination, the movement of particles and bubbles across the backlight generates distinct positive and negative events. The leading interface of particle/bubble blocks the backlight, generating negative events (shown in blue in **Figure. 3**). When the trailing interface passes, and the light becomes visible again, positive events shown in white are generated. Extending the accumulation time causes these events to form streak patterns along the direction of motion. Since the newly generated events are layered over older ones, the positive events from the trailing interface partially overwrite the earlier negative events. Therefore, the remaining blue pixels represent the object region between the leading and trailing interfaces. To fully represent the shadows of bubbles or particles, the selected accumulation time should exceed the ratio of bubble or particle size to velocity.

A distinct feature of the bubbles is the light reflection spots, which generate positive events under pulsed illumination due to insufficient blockage of the backlight. Under continuous illumination, these spots correspond to localized increases in intensity, surrounded by negative events that form the bubble shadow.

Following the raw video extraction, image correction (raw to real-world) was performed. Cusing DaVis and subsequently processed in ImageJ to mitigate background and interface noise. Under pulsed illumination, the reconstructed frames are binary in nature, whereas continuous illumination produces frames comprising three event states: positive events (white), negative events (blue), and inactive pixels corresponding to no events (black).

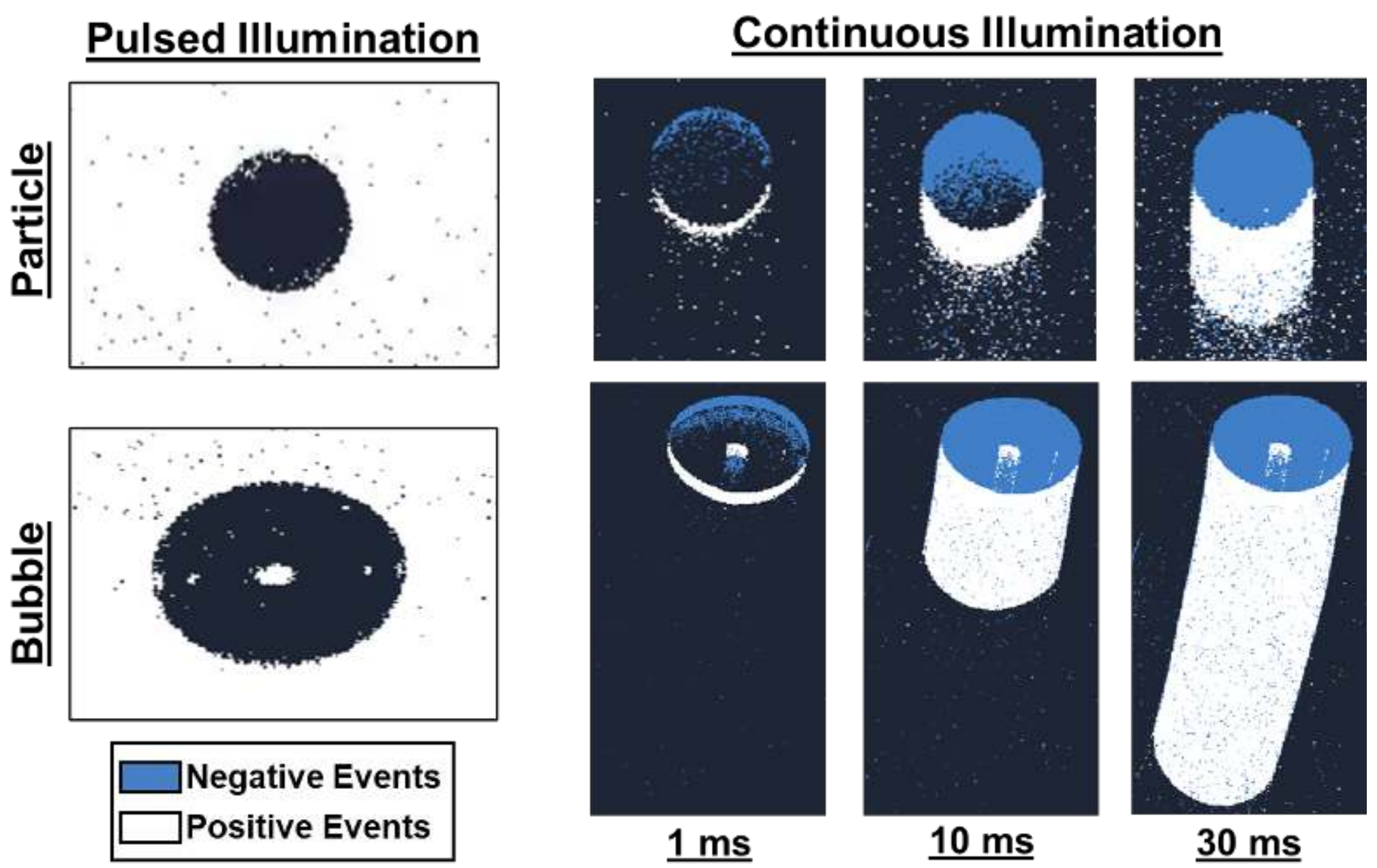

**Figure. 3** EVS camera frames of a particle and bubble under pulsed (left) and continuous (right) illumination. Continuous illumination results are shown for accumulation times of 1 ms, 10 ms, and 30 ms. Blue, white, and black pixels represent decreases in light intensity, increases in light intensity, and inactive pixels, respectively.

To track the particles and bubbles within the three-dimensional measurement domain, a custom MATLAB processing framework, shown in **Figure. 4** was developed. The framework utilizes the intrinsic and extrinsic calibration parameters, obtained from the DaVis pinhole camera calibration model, to generate the projection matrices for each EVS camera's viewing angle. These projection matrices were subsequently used to establish the geometric relationship between the three-dimensional global coordinate system and the corresponding two-dimensional image planes during epipolar matching and triangulation.

The three-dimensional positions of detected particles and bubbles were estimated using a multi-view reconstruction framework. Each camera view was independently segmented to extract candidate regions. To address segmentation inconsistencies and merged detections in cases of overlap, a cross-view-based target object count per frame was defined as the maximum number of discrete regions identified across the three synchronized cameras.

In cases of discrepancy, large-area regions were identified as potential merged detections and subjected to morphological decomposition. These regions were analyzed by comparison with the median bubble size computed across all frames, enabling estimation of the number of constituent bubbles within each cluster. Merged regions were further processed using a Euclidean Distance Transform (EDT) (Maurer et al., 2003) to obtain a spatial distance field, where pixel intensities correspond to distances from the nearest boundary. To enhance robustness, the distance maps were smoothed using a Gaussian filter, facilitating stable identification of local maxima corresponding to bubble centroids.

To improve reconstruction robustness during bubble overlapping, pairwise epipolar consistency constraints were introduced between all synchronized camera combinations. For each camera pair, epipolar-distance cost matrices were constructed and optimized using Hungarian assignment, enabling reliable and globally consistent correspondence matching across views. When detections were temporarily unavailable in one camera, a two-camera reconstruction fallback was used to directly recover the three-dimensional position from the remaining synchronized views. This strategy helped maintain continuous trajectories during periods of temporary visibility loss and severe projection overlap.

A distinguishing feature of the bubble tracking framework is the utilization of reflection spots for both trajectory refinement and independent validation. While silhouette-based tracking provides complete morphological information of the bubble, bright reflection points generated under pulsed illumination are simultaneously extracted. This dual-mode formulation facilitates validation by cross-referencing trajectories derived from silhouette centroids with those obtained from reflection spots. This approach is particularly advantageous in high-density regions, where bubble silhouettes merge into complex structures, whereas reflection spots typically remain spatially isolated.

The spatial coordinates were reconstructed using a Direct Linear Transformation (DLT) formulation (Abdel-Aziz & Karara, 2015). For each matched centroid across camera views, a system of linear equations was constructed based on the projection model. The resulting overdetermined system was solved based on a least-squares approach using Singular Value Decomposition (SVD) (Golub & Reinsch, 1971), thereby accounting for pixel-level noise and calibration uncertainty. The homogeneous coordinates were subsequently normalized to yield Cartesian coordinates in the global reference frame.

A physics-constrained temporal tracking framework was developed to reconstruct continuous trajectories of particles and bubbles from sequential three-dimensional position estimates. Centroid linking was performed using a Lagrangian nearest-neighbour strategy constrained by a maximum physically admissible displacement per frame, derived from an upper-bound velocity and the camera frame rate. A temporal coherence constraint was additionally imposed to maintain trajectory continuity in the presence of intermittent detection loss by limiting associations to a maximum allowable frame gap.

To further improve temporal trajectory association, a predictive kinematic tracking framework was introduced. Particle velocities were estimated using least-squares regression over recent trajectory histories, thereby enabling the prediction of expected particle positions between consecutive frames. Candidate associations were then evaluated using a combined cost metric that accounted for spatial proximity, velocity consistency, and agreement in motion direction. This approach improved robustness in situations involving ambiguous particle interactions and temporary trajectory crossings.

To resolve ambiguous associations, a conflict-resolution mechanism was introduced. When multiple candidate points were assigned to the same trajectory within a single frame, selection was based on consistency with the recent motion history. Each candidate was evaluated according to its deviation from the preceding trajectory segment, and the most kinematically consistent observation was retained. Remaining candidates were either used to initiate new trajectories or discarded if they did not persist for a few frames. Finally, instantaneous velocity and acceleration were computed using finite-difference approximations along each trajectory, enabling time-resolved kinematic analysis of bubble and particle motion.

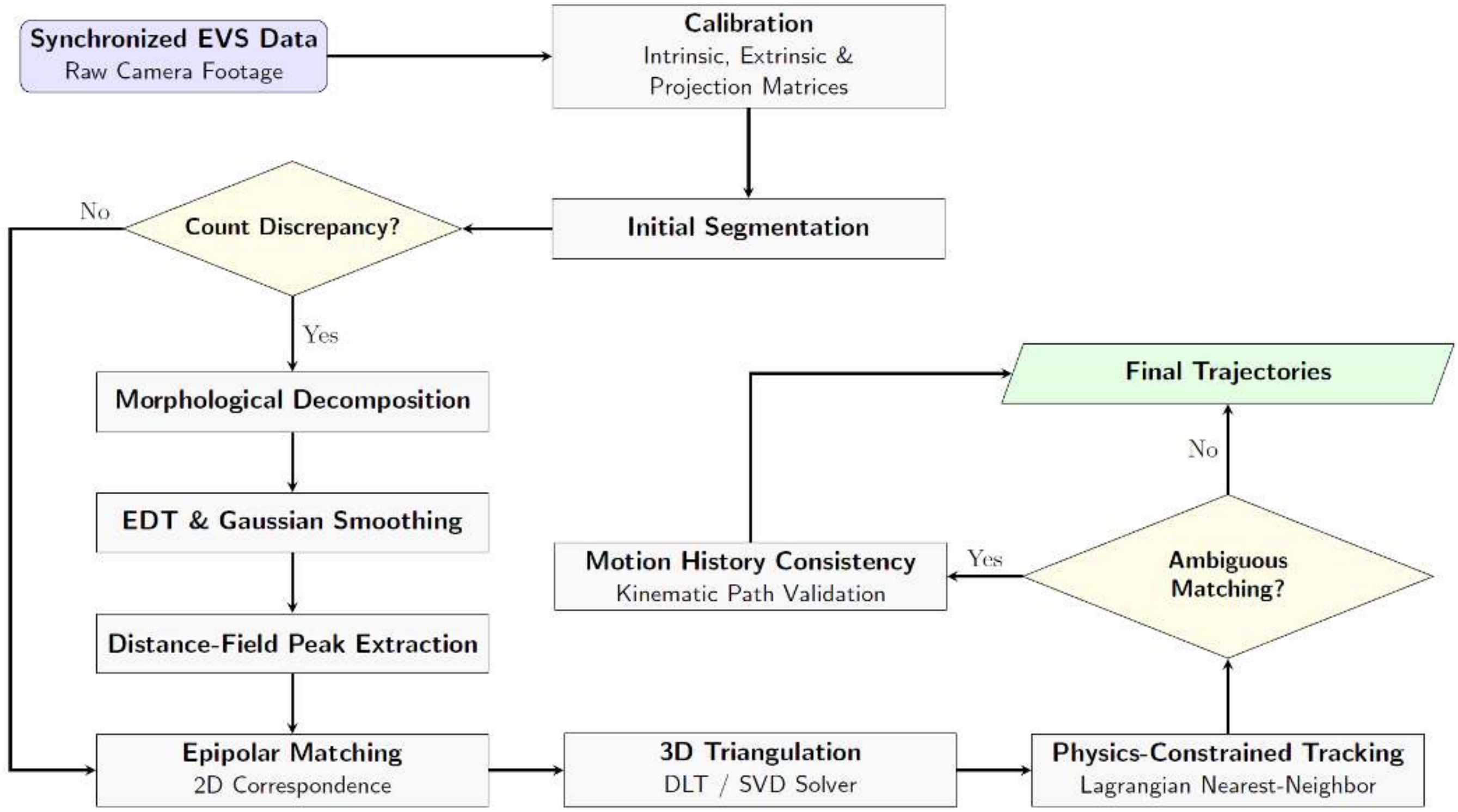

**Figure. 4** Overview of the multi-view reconstruction and physics-constrained tracking framework for three-dimensional bubble and particle motion analysis. The method integrates segmentation correction, epipolar matching, DLT triangulation, and motion-history-based trajectory validation.

Streaks produced under continuous illumination provide important supplementary kinematic information for trajectory tracking. Similar to bubble or particle shadow centroids, the centroid of the streak  can also be tracked, as it closely follows the motion path of the shadow. More significantly, the absolute streak length is directly proportional to the particle velocity, as described by the following relationship:

$$Streak\ Velocity\ (mm/s) = \frac{Streak\ Length\ (mm)}{Accumulation\ Time\ (s)} \tag{1}$$

Measuring streak length requires isolating the interface between the shadow cap and the trailing streak region. The algorithm analyses the left and right streak boundaries on a pixel-by-pixel basis, estimating local slopes through a moving-window approach to determine the trajectory vector. The enclosed region between these boundaries is then examined to identify the bottom interface of the streak. Because the trailing edge is particularly sensitive to optical background noise, a second-degree polynomial is fitted to the lowest detected pixels, providing a stable mathematical description of the streak tail. The representative 2D streak length is subsequently defined as the vector extending from the midpoint of the top interface to the point where it intersects the fitted polynomial boundary. These extracted geometric features are illustrated in **Figure. 5**. The corresponding 2D measurements from each synchronized camera view are then triangulated to reconstruct the absolute three-dimensional displacement and velocity.

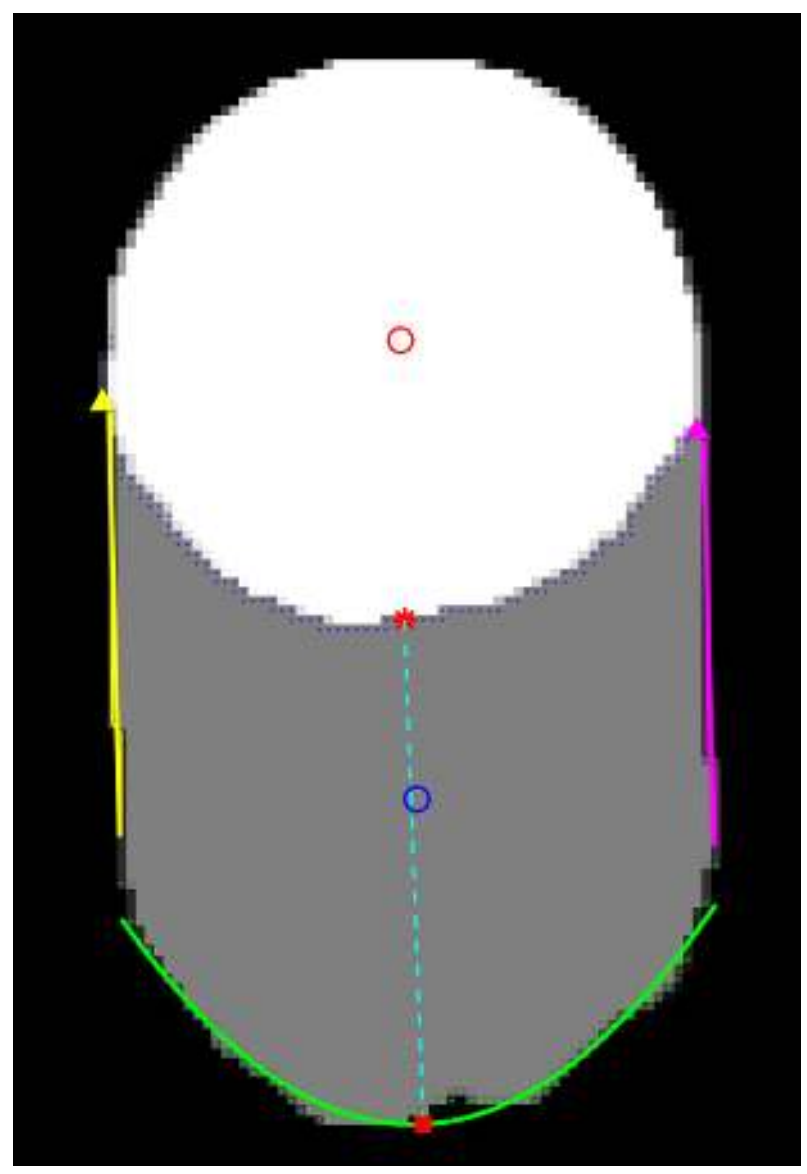

**Figure. 5** Two-dimensional image processing for the extraction of particle streak features from footage acquired under continuous illumination.

To benchmark and validate the tracking framework, a series of synthetic rendering cases were generated to provide a numerical ground-truth dataset. Three-dimensional particle trajectories were analytically prescribed and projected onto each camera plane using the same intrinsic and extrinsic calibration parameters obtained from the experimental calibration procedure. Synthetic particle images were subsequently rendered for each synchronized optical view, enabling direct evaluation of epipolar matching performance, triangulation accuracy, and temporal trajectory reconstruction under controlled motion conditions.

The validation cases were constructed with progressively increasing kinematic complexity, ranging from linear trajectories to projection-crossing paths, spiral motion, and staggered triangular-array releases. The method achieved a global root mean square error (RMSE) of 0.224 mm, 0.3624 mm, and 0.015 mm across the respective test cases, as illustrated in **Figure. 6**. These scenarios emphasized trajectories that overlapped in individual two-dimensional camera projections while remaining spatially separated in three-dimensional space. These cases specifically validated the robustness of the Direct Linear Transformation (DLT) reconstruction and motion-history-based tracking framework in resolving geometric ambiguities.

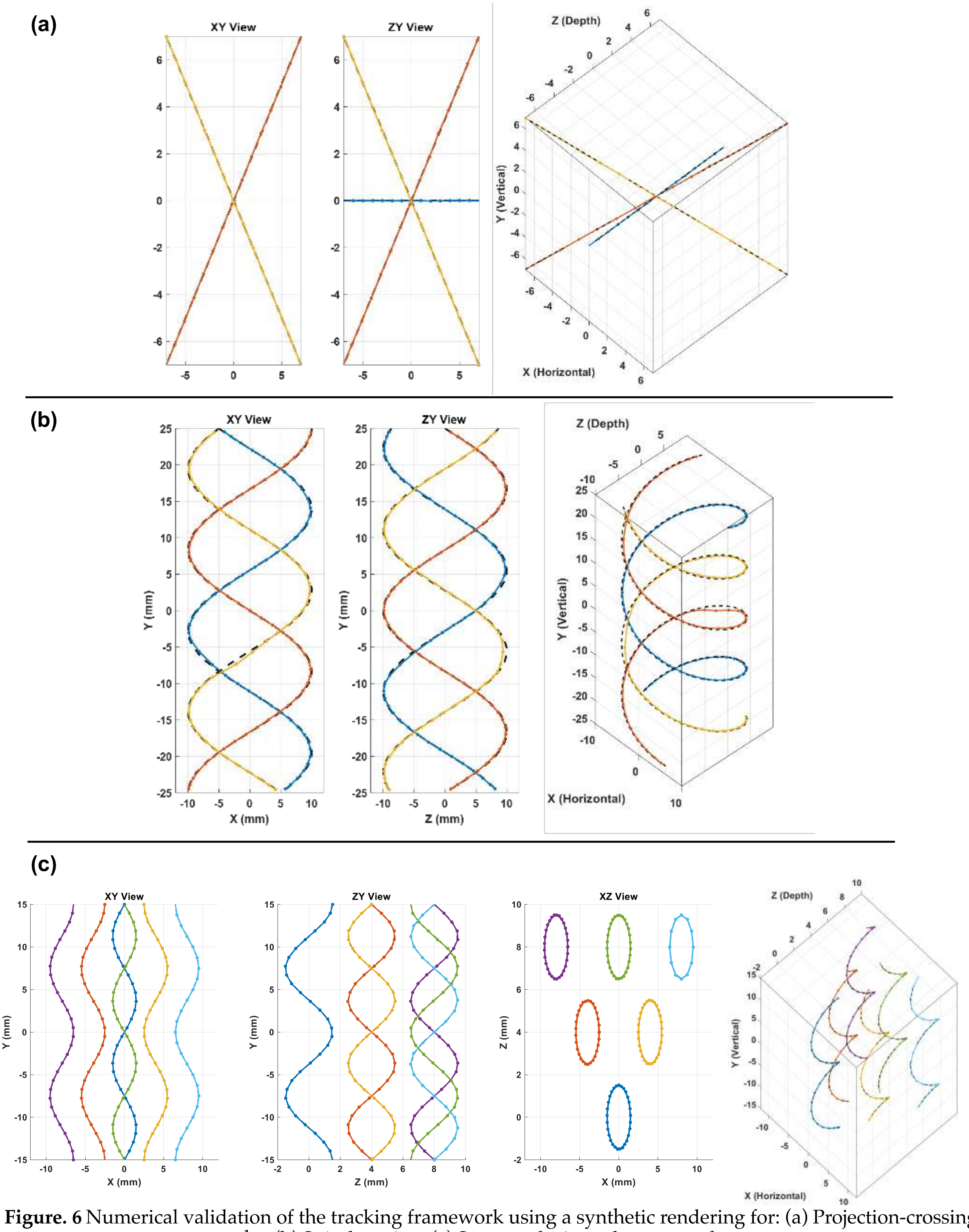

**Figure. 6** Numerical validation of the tracking framework using a synthetic rendering for: (a) Projection-crossing paths, (b) Spiral motion, (c) Staggered triangular-array releases.

## 4. Results and Discussion

The experimental configuration and tracking algorithms were first evaluated in capturing particle release from the bottom of the tank following rotation of the motorized gate from three different chambers. Ideally, particle release would occur without interference. However, the initial gate opening dynamics and particle adhesion to the gate were found to significantly influence both the release behavior and the resulting trajectories. Furthermore, the disparity between chamber and particle diameters affected the release process by causing particles to move along the gate as it rotated to open. In addition, some particles were in direct contact with the chamber walls, leading to frictional interactions. Finally, the wake generated by leading particles was observed to influence the motion of subsequent particles during the release process.

The obtained trajectory results under pulsed illumination in **Figure. 7** reveals that particle motion is predominantly vertical across all chambers. Horizontal (x-direction) and depth variations are comparatively small and are primarily associated with transient release effects and wake-induced interactions. Despite these perturbations, the overall motion remains strongly aligned with positive buoyancy settling. Quantitatively, the particles exhibit a characteristic velocity of the order of 70 mm/s, with minor variations observed between individual trajectories. This value represents the average translational speed during the post-release motion phase and is consistent across the three chamber configurations within experimental uncertainty.

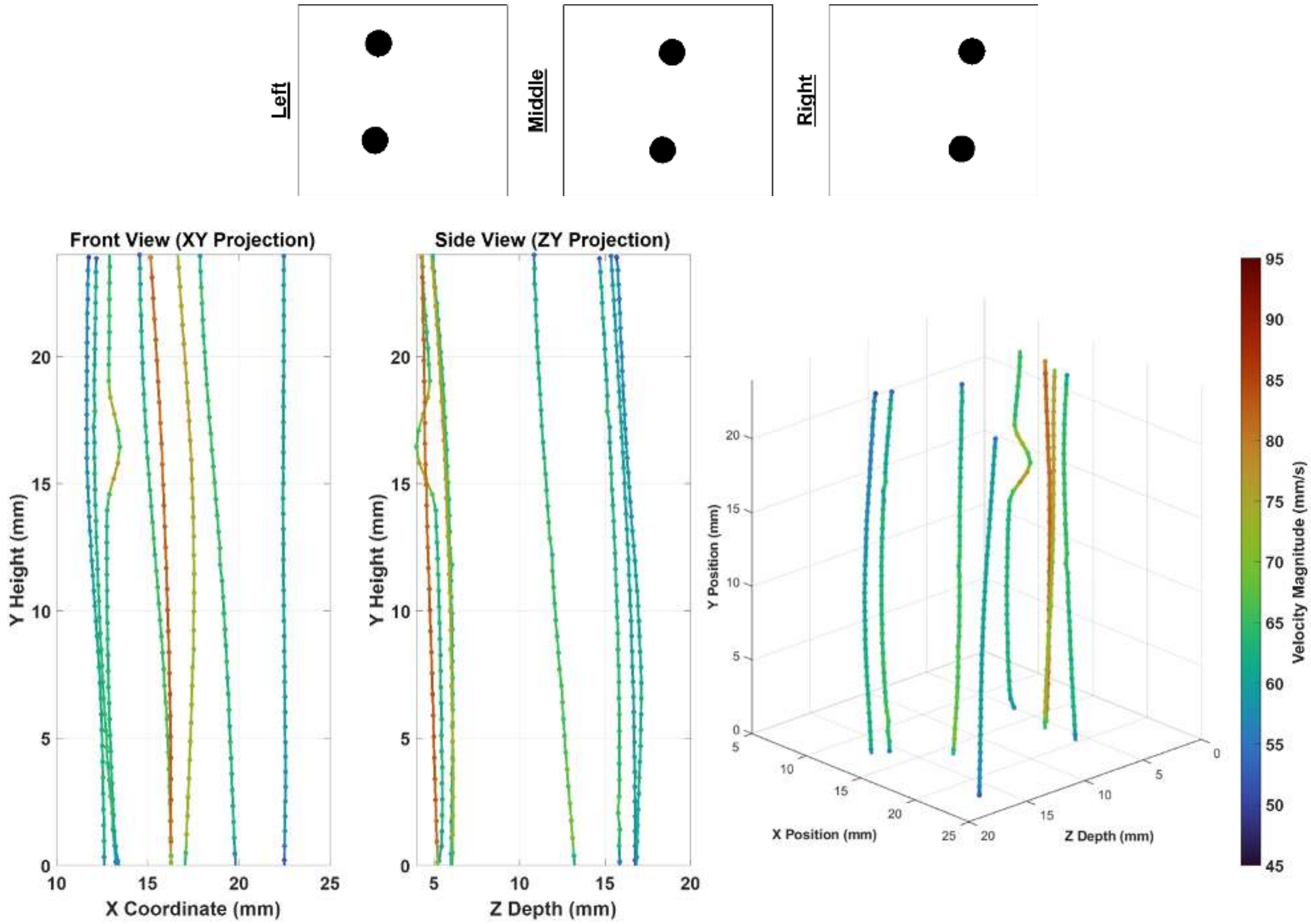

**Figure. 7** Example trajectories of particles following the release from a gated particle chamber under pulsed illumination.

Particle release tracking using continuous back illumination was also performed, as shown in **Figure.** *8*, where both particle shadows and corresponding streaks were utilized for trajectory reconstruction. The differences between tracks derived from shadow-based centroid tracking and streak-based analysis were minor and primarily attributed to noise at the lower interface of the streak profiles. Estimation of instantaneous velocity and direction from the three-dimensional streak lengths showed strong agreement with the discrete, frame-by-frame centroid tracking method. The mean velocity comparison indicates an overall agreement of 93.5% across all tracked particles. A small but consistent systematic deviation was observed, with the streak-based method yielding velocities approximately 4.5% higher than those obtained from discrete centroid tracking. This bias suggests that discrete temporal sampling tends to smooth or linearize higher-frequency fluctuations in the trajectory, whereas streak-based measurements capture continuous motion more directly. The mean orientation angles also demonstrated strong consistency between the two approaches, with centroid-based tracking yielding an average angle of −1.80°, compared to −1.15° obtained from streak-based measurements. Overall, both methods show excellent agreement, with minor deviations attributed to temporal discretization effects and streak-edge noise.

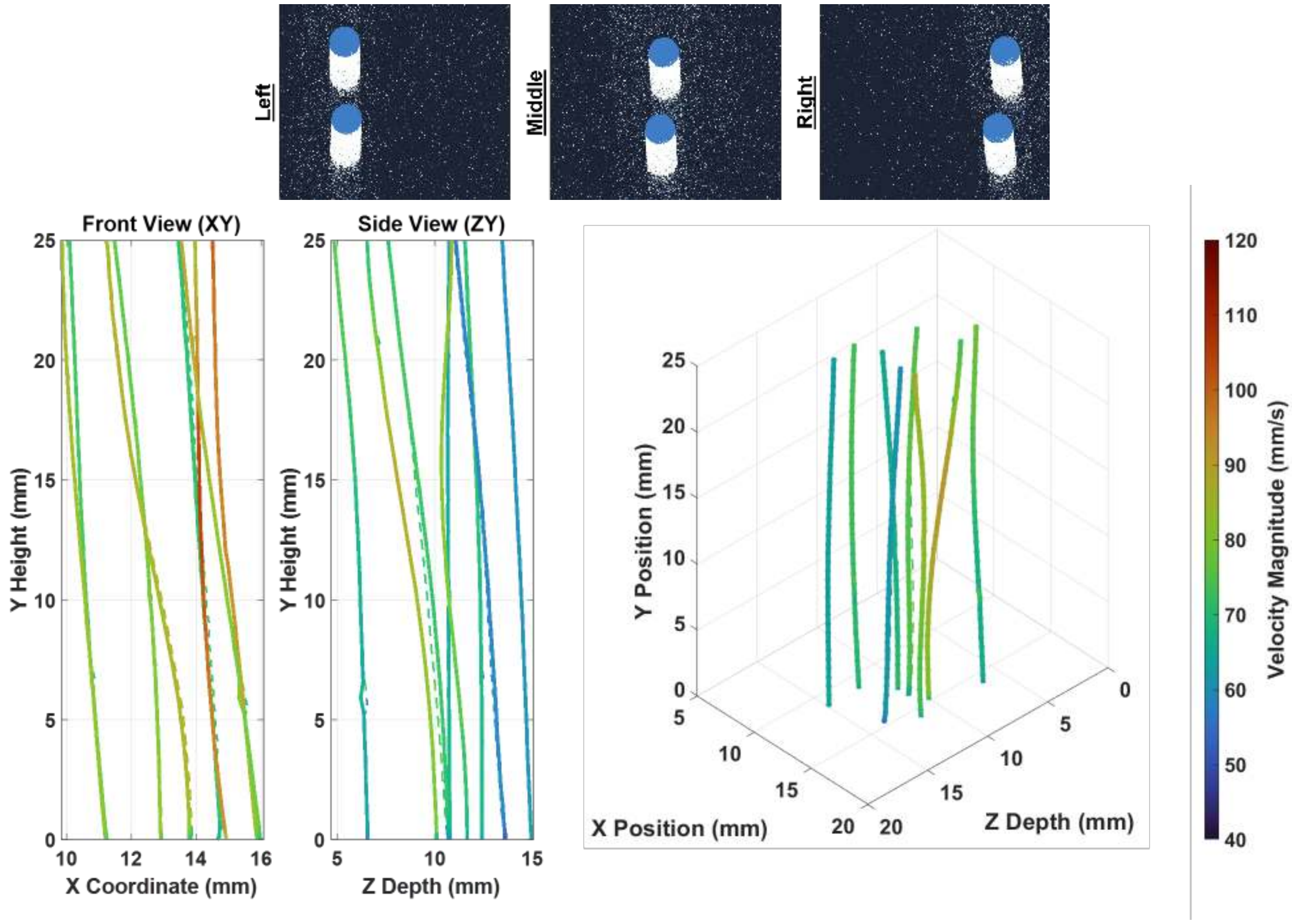

**Figure. 8** Example trajectories of particles following the release from the particle chamber with continuous illumination (the dashed line represents the tracks created by following the streak centroid).

While particle release through the chamber provided a suitable test case for evaluating the three-camera system and tracking algorithms, limited experimental repeatability motivated the development of an alternative release mechanism. As shown in **Figure. 9**, this limitation was addressed by introducing a centrally positioned claw mechanism within the tank, which releases particles upon plunging. The resulting trajectories exhibit predominantly vertical motion, indicating that the 3D triangulation and motion reconstruction were successfully implemented with good accuracy. Furthermore, the reconstructed velocity profiles reveal a smooth and physically consistent transient acceleration phase immediately following particle release from rest.

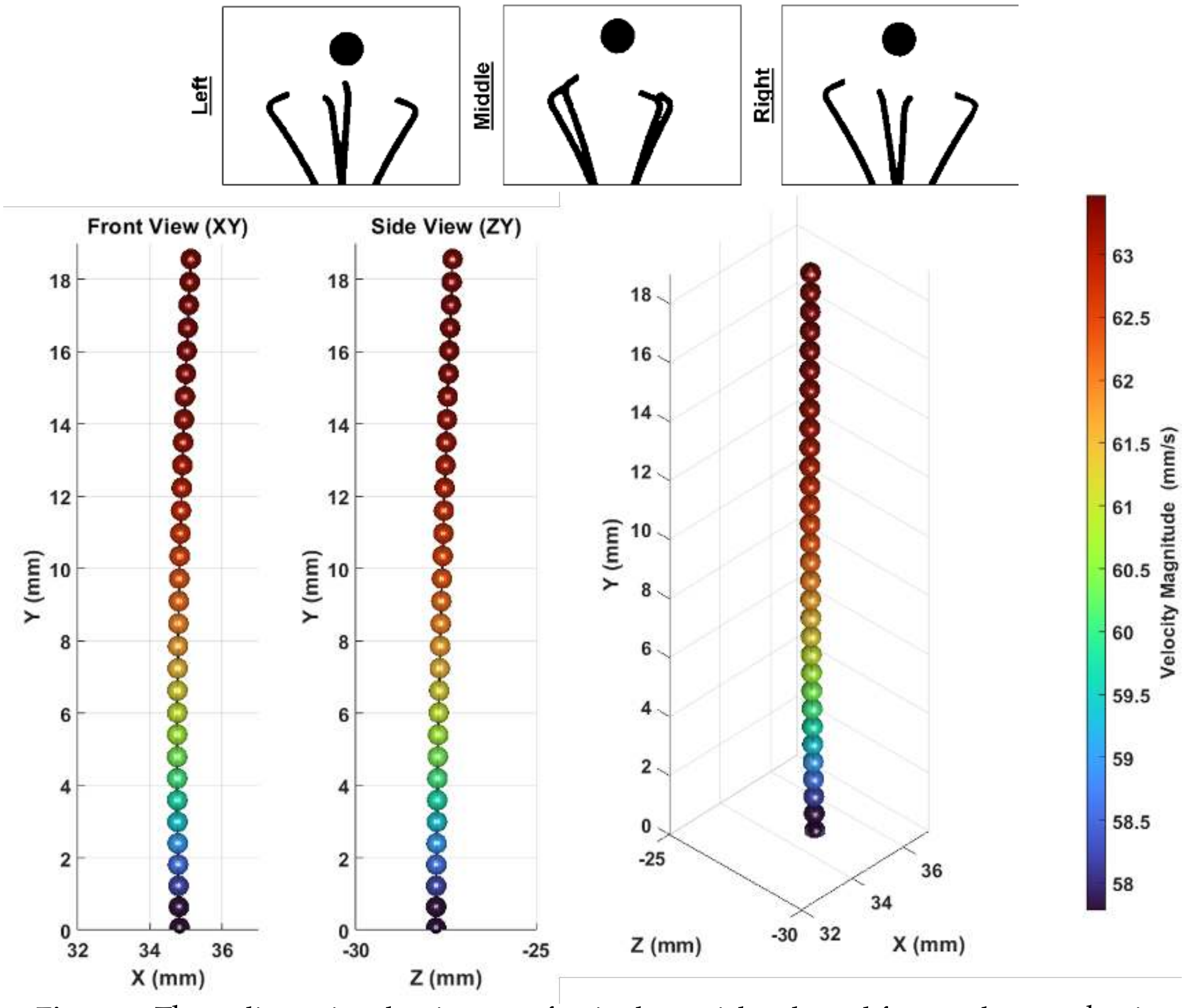

**Figure. 9** Three-dimensional trajectory of a single particle released from a claw mechanism.

Following the analysis of the precisely manufactured particles, bubble plume release from the bottom of the tank was investigated. Bubble plumes were generated using compressed gas supply flow rates of 10 mL/min and 20 mL/min, corresponding to bubble generation frequencies of approximately 16 Hz and 45 Hz, respectively. Bubble motion under both flow conditions was recorded using pulsed and continuous illumination. However, under the higher gas supply condition, bubble streaks were frequently obscured by subsequently released bubbles owing to wake-induced interactions. Consequently, to better demonstrate the measurement capabilities of both illumination approaches, continuous illumination was employed under the lower gas flow condition, where bubble trajectories remained comparatively unobstructed. This enabled a direct evaluation of event-streak-based measurements for instantaneous velocity estimation alongside conventional spatiotemporal bubble centroid tracking.

The reconstructed three-dimensional trajectories of the bubble plume at a 16 Hz injection frequency and a 10 ml/min gas flow rate reveal distinct structural regions as the bubbles ascend, as shown in **Figure. 10**. At this low supply rate, the bubbles are only generated through the central inlet, which is reflected in the trajectories between Y = 0 to Y = 5mm. These trajectories are localized and dense, corresponding to the primary core of the bubble plume immediately following detachment. As the bubble rises, significant lateral dispersion is observed, resulting in a characteristic conical plume expansion. This expansion could be explained by the bubble wake interaction (Huang et al., 2025) as well as the impurities in the water, and the uneven distribution of surfactant molecules along the bubble interface (Tagawa et al., 2014).

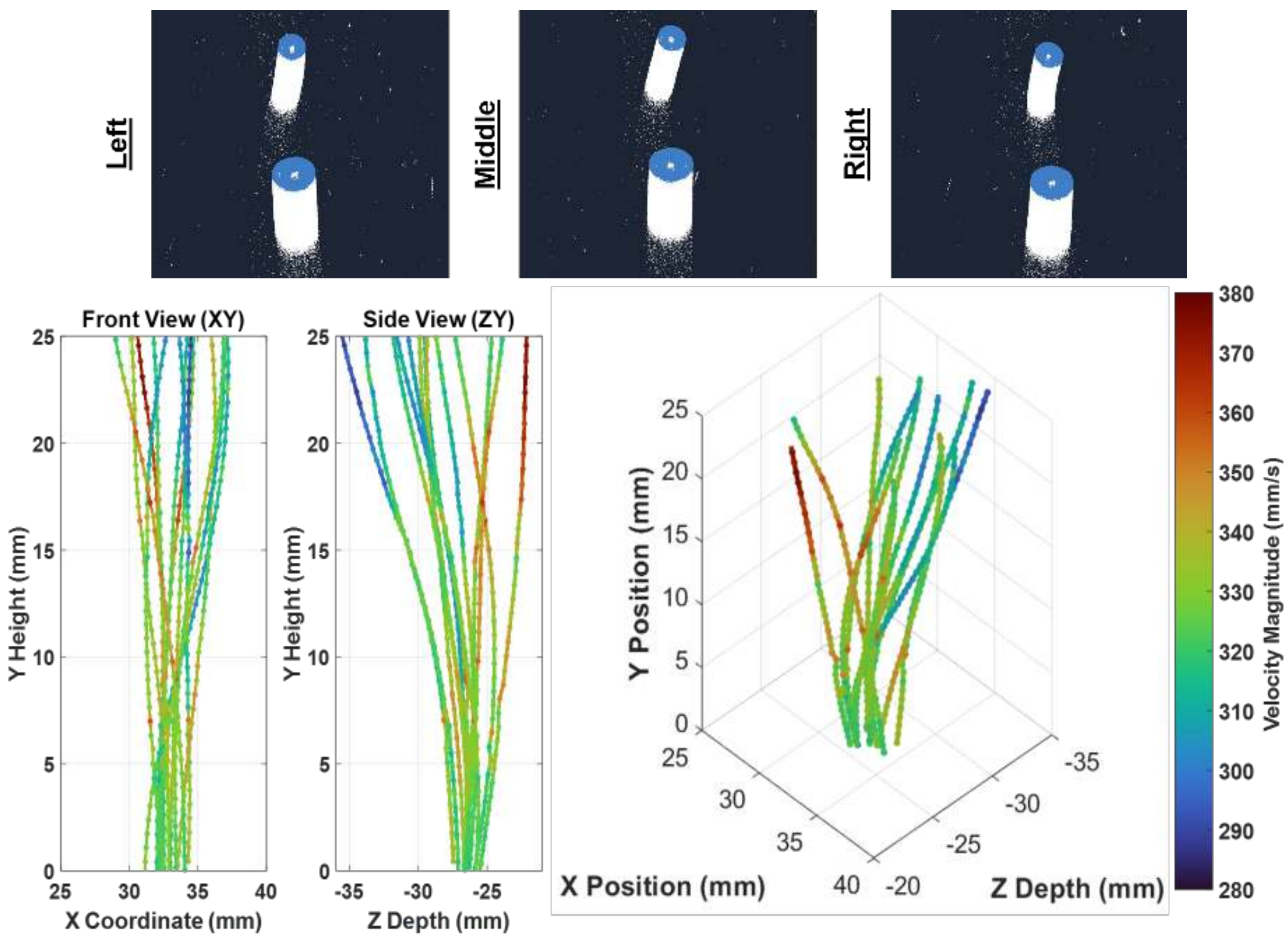

**Figure. 10** Three-dimensional trajectories of a bubble plume generated from a supplied gas with a flow rate of 10 mL/min under continuous illumination.

The velocity magnitude profiles mapped along the reconstructed trajectories provide direct insight into momentum transfer and hydrodynamic interactions within the plume. At the release point, bubbles enter the domain with initial velocities in the range of 310–330 mm/s. As buoyancy forces progressively overcome initial inertial effects, a distinct acceleration phase is observed. A clear path-dependent velocity structure emerges within the plume topology. This behavior provides direct evidence of bubble–wake interaction, whereby trailing bubbles moving through the upward-inducing, low-drag wake of leading bubbles experience reduced hydrodynamic resistance, enabling them to exceed the terminal velocity of an isolated bubble.

Velocities determined through streak-length measurements were compared to those obtained from bubble centroid tracking and exhibited strong agreement, yielding an overall velocity consistency exceeding 97% between the two independent methodologies. On a frame-by-frame basis, the local velocity discrepancy remains low, typically between 1% and 5% (with an average relative deviation of approximately 2.5%), accompanied by an average angular discrepancy of less than 5%.

To further evaluate the tracking framework under increased multi-inlet bubble densities and intensified hydrodynamic interactions under pulsed illumination, the gas flow rate was increased to 20 mL/min, yielding a bubble generation frequency of approximately 45 Hz across the two inlets. The reconstructed three-dimensional trajectories for this high-frequency regime, shown in **Figure. 11**, exhibit a markedly different spatiotemporal structure compared to the lower-frequency 16 Hz condition, accompanied by substantial bubble overlap within the camera views.

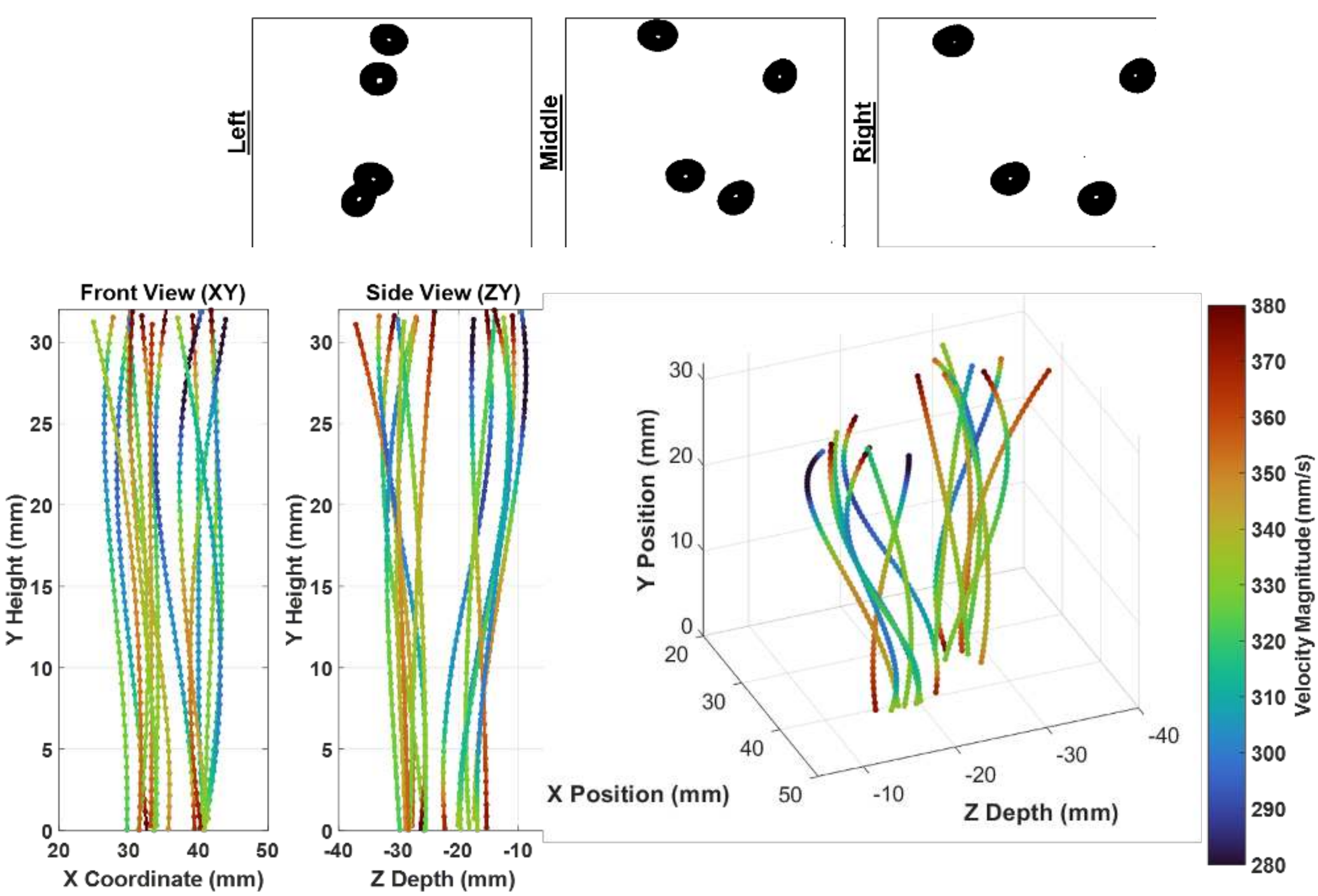

**Figure. 11** Three-dimensional trajectories of a bubble plume generated from a supplied gas with a flow rate of 20 mL/min under pulsed illumination.

The most prominent feature is the significant lateral expansion of the plume. Although bubble detachment remains spatially concentrated near the nozzle exit, the plume rapidly diverges during ascent, reaching approximate spreads of 25 mm along the X-axis and 30 mm along the Z-axis. The elevated injection frequency intensifies wake-induced hydrodynamic interactions, producing a highly non-uniform velocity distribution throughout the plume. The velocity field reveals distinct high-velocity drafting regions coexisting with slower peripheral trajectories. Bubbles propagating within the wakes of preceding bubbles experience reduced hydrodynamic drag and accelerated ascent, whereas bubbles displaced from the central wake structures exhibit comparatively lower rise velocities. The resulting velocity stratification provides clear evidence of intensified wake-drafting behaviour under high-frequency injection conditions.

From a tracking and instrumentation standpoint, the 45 Hz regime constitutes a rigorous validation case for the multi-view three-dimensional tracking framework under pulsed illumination. The dense front and side projections exhibit substantial visual overlap and highly intersecting trajectories. Nevertheless, continuous and independent trajectory identities are maintained throughout the measurement domain, demonstrating the robustness of asynchronous event-based sensing for high-density multiphase flow characterization.

While EVS cameras have demonstrated strong capability for high-temporal-resolution bubbly flow measurements at reduced data rates and comparatively low cost, several critical limitations must be considered when deploying synchronized multi-camera EVS systems. The principal limitation remains event oversaturation. In contrast to HS cameras, which acquire full image frames at fixed sampling rates, EVS camera pixels operate asynchronously by independently detecting temporal variations in light intensity. These events are subsequently read and time-stamped by the row arbiter architecture. When the event rate exceeds the sensor saturation threshold, undesirable phenomena arise, including event loss, temporal distortion, and timestamp inaccuracies. These timing inconsistencies are particularly problematic in synchronized multi-camera configurations, where desynchronization of a single camera during acquisition may render the complete multi-view dataset unusable.

To prevent event oversaturation, the event rate must be carefully controlled. This rate depends on both the number of activated pixels and how frequently those pixels generate events. Under continuous illumination, the number of activated pixels is determined by the number of bubbles or particles present within the Region of Interest (ROI) and their pixel size in the frame, while the activation frequency is primarily governed by particles or bubbles' velocities. Under pulsed illumination, the activated pixels mainly correspond to illuminated background regions produced during each light pulse, excluding the shadowed regions created by particles and bubbles. In this case, the activation frequency increases directly with the LED illumination pulse rate. Pulse illumination offers improved controllability through the appropriate selection of pulse frequency and ROI, ensuring that the EVS camera event generation rate remains below the event saturation threshold. However, based on observations with the utilized EVS camera model, sporadic spikes in event rate were observed. The most notable anomaly was event loss due to sensor oversaturation during calibration target acquisition near the midpoint, as illustrated in **Figure. 12**. Restarting the camera acquisition was sufficient to resolve this issue. However, if undetected, this problem could have severely compromised the calibration dataset or rendered it unusable.

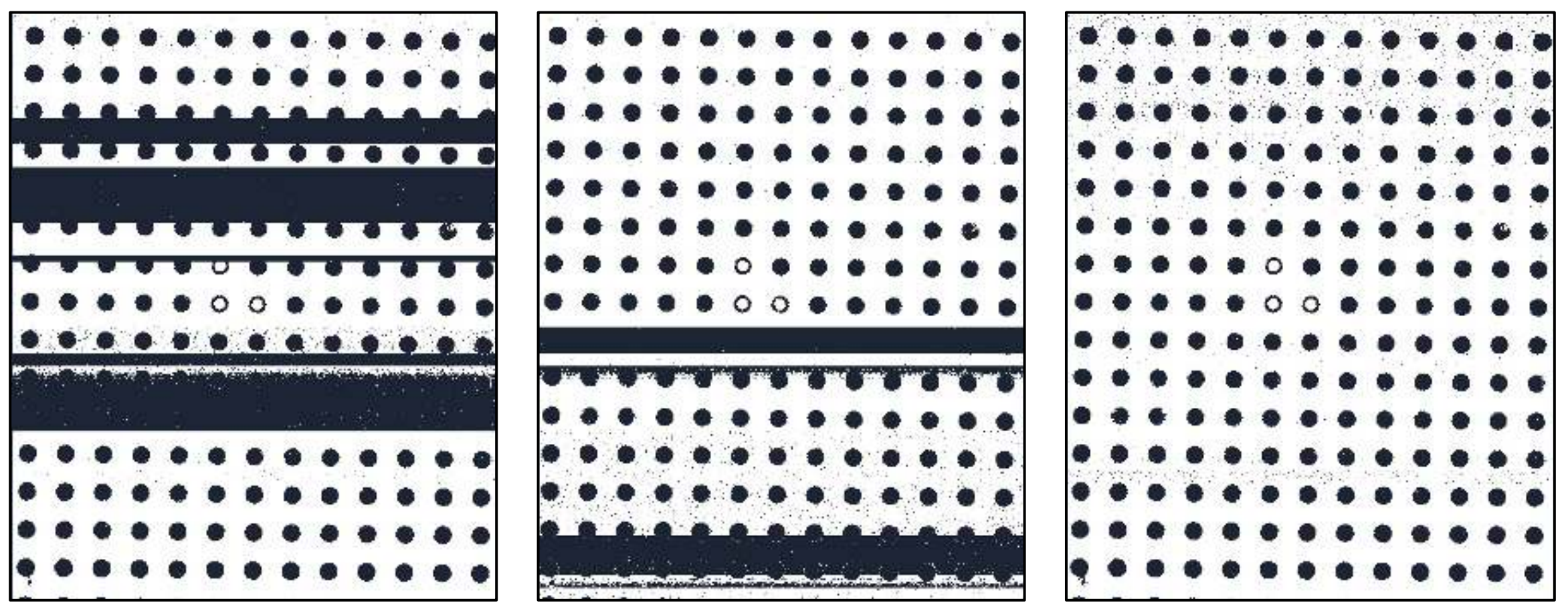

**Figure. 12** Missing event due to oversaturation during calibration target image acquisition

## 4. Conclusions

This study successfully demonstrated a multi-EVS camera framework for three-dimensional tracking and kinematic characterization of particles and bubble plumes. The experimental configuration consisted of an octagonal tank equipped with both a particle release mechanism and an air diffuser for controlled bubble generation. Camera synchronization and pulsed LED illumination were achieved using a signal generator and dedicated driver circuitry, while camera calibration was performed using pulsed-illumination recordings of a target acquired at multiple depths.

A comprehensive in-house computational framework was developed for image processing and three-dimensional motion trajectory reconstruction. When benchmarked against synthetic rendering cases of increasing kinematic complexity, the framework achieved sub-millimeter global accuracy, with RMSE values ranging from 0.015 to 0.362 mm. In physical experiments, the EVS framework successfully resolved dense multiphase interactions, producing smooth and physically consistent 3D trajectories for both particle releases and high-frequency (16–60 Hz) bubble plumes. Quantitative evaluation showed an overall velocity consistency exceeding 97% between conventional centroid tracking and independent velocity estimates derived from continuous-illumination event streaks, with local frame-by-frame deviations averaging approximately 2.5%.

Overall, the presented methodology establishes a scalable framework for high-speed volumetric tracking and multiphase flow characterization. The integration of advanced image processing, temporal trajectory association, and 3D reconstruction provides a robust foundation for future investigations involving dense bubbly flows and complex particle interactions. Although EVS cameras offer distinct advantages, including real-time sensing, extended recording duration at low data rates, and relatively low system cost, event oversaturation remains a critical hardware limitation affecting measurement reliability in ultra-dense regimes. Continued advancements in event sensor technology are expected to mitigate this limitation in the future.

## Acknowledgments

Kuppuraj Rajamanickam would like to acknowledge the funding support received from the Marie Curie Fellowship through UKRI – Horizon Europe guarantee funding scheme (EP/X026248/1)